\documentclass[a4paper,bibtotoc]{scrartcl}
\usepackage{amsmath}
\usepackage{amssymb}
\usepackage[latin9]{inputenc}
\usepackage{amsthm}
\usepackage{setspace}
\usepackage{longtable}

\begin{document}
\theoremstyle{definition}
\newtheorem{theorem}{Theorem}[section]
\newtheorem{definition}[theorem]{Definition}
\newtheorem{corollary}[theorem]{Corollary}

\newcommand{\cfr}{cf_{\{0\}}^{\mathbb{R}}}
\newcommand{\lev}{\operatorname{lev}}
\newcommand{\Lev}{\operatorname{Lev}}
\newcommand{\bas}{\operatorname{bas}}
\newcommand{\dom}{\operatorname{dom}}
\newcommand{\num}{\operatorname{num}}
\newcommand{\cfrn}[1]{\langle \cfr \rangle^{#1}}

\author{Arno Pauly\thanks{Email: arno.pauly@cl.cam.ac.uk} \\ University of Cambridge}
\title{On the (semi)lattices induced by continuous reducibilities}

\maketitle

\abstract{Continuous reducibilities are a proven tool in computable analysis, and have applications in other fields such as constructive mathematics or reverse mathematics. We study the order-theoretic properties of several variants of the two most important definitions, and especially introduce suprema for them. The suprema are shown to commutate with several characteristic numbers.}

\section{Introduction}
Studying discontinuity of functions is an interesting topic on its own, an observation that is fortified by noting that continuity behaves similar to computability in the framework of computable analysis. This suggests to compare the discontinuity of functions through continuous reducibilities. In the present paper, a continuous version of bounded Turing reducibility ($\leq_2$) and a continuous version of many-one reducibility ($\leq_0$) will be studied.

Another motivation for the study of these relations stems from parallels between computable analysis and Bishop's constructive mathematics (\cite{bishop}) for $\leq_0$ and $\leq_2$ and between computable analysis and reverse mathematics (\cite{simpson2}) for $\leq_2$. As spelled out in \cite{troelstra}, statements of the form $f \leq_0 g$ often correspond to set inclusions in constructive mathematics. The relationship between discontinuity and inconstructibility was studied in \cite{weihrauchc}. In reverse mathematics, $f \leq_2 g$ can correspond to the observation that a statement $A$ can be proven with no more axioms than needed for proving $B$, as demonstrated in \cite{gherardi}. Neither of the two parallels seems to be strict, but both were successfully used to derive new insight in one of the respective fields.
\section{Preliminaries}
\subsection{Topology}
Given a set $X$, a topology $\mathcal{T}$ on $X$ is a set of subsets of $X$ including $\emptyset$ and $X$, which is closed under formation of arbitrary unions and countable intersections. The elements of a topology are called open sets, their complements are called closed sets. Since any union of open sets returns an open set, any intersection of closed sets is closed, enabling the definition of $cl U$ as the smallest closed set containing $U \subseteq X$. For each set $X$, the discrete topology is given by $\mathcal{T}_d = \{U \subseteq X\}$ and the indiscrete topology is given by $\mathcal{T}_i = \{\emptyset, X\}$.

A topological space is a set equipped with a topology. Given a set-indexed family $(X_i, \mathcal{T}_i)_{i \in I}$, the coproduct $\coprod \limits_{i \in I} (X_i, \mathcal{T}_i)$ is the set $\bigcup \limits_{i \in I} \{i\} \times X_i$, equipped with the smallest topology $\mathcal{T}$ satisfying $\{(\{i\} \times U) \mid U \in \mathcal{T}_i\} \subseteq \mathcal{T}$ for all $i \in I$. The product $\prod \limits_{i \in I} (X_i, \mathcal{T}_i)$ is the set $\prod \limits_{i \in I} X_i$ equipped with the smallest topology containing  $\{\prod \limits_{i \in I} U_i \mid \forall i \in I \ U_i \in \mathcal{T}_i, |\{ i \in I \mid U_i \neq X_i\}| < |\mathbb{N}|\}$. For a topological space $(X, \mathcal{T})$ and a subset $Y \subseteq X$, a topology on $Y$ is defined as $\mathcal{T}_Y = \{U \cap Y \mid U \in \mathcal{T}\}$.

A function $f$ between topological spaces $(X, \mathcal{T})$ and $(Y, \mathcal{S})$ is a function $f: X \to Y$. It is continuous, if it satisfies $f^{-1}(U) \in \mathcal{T}$ for all $U \in \mathcal{S}$. The function  $\iota_j: (X_j, \mathcal{T}_j) \to \coprod \limits_{i \in I} (X_i, \mathcal{T}_i)$ defined through $\iota_j(x) = (j, x)$ is continuous, as well as the function $\pi_j : \prod \limits_{i \in I} (X_i, \mathcal{T}_i) \to (X_j, \mathcal{T}_j)$ that is the projection to the $j$th entry. The inclusion of $(Y, \mathcal{T}_Y)$ in $(X, \mathcal{T})$ for $Y \subseteq X$ is also continuous. If $f: (X, \mathcal{T}) \to (Y, \mathcal{S})$ is continuous, and $Z \subseteq X$, so is $f_{|Z}: (Z, \mathcal{T}_Z) \to (Y, \mathcal{S})$.

For a family of continuous functions $(f_i : (X_i, \mathcal{T}_i) \to (Y_i, \mathcal{S}_i))_{i \in I}$, the function  $\coprod \limits_{i \in I} f_i: \coprod \limits_{i \in I} (X_i, \mathcal{T}_i) \to \coprod \limits_{i \in I} (Y_i, \mathcal{S}_i)$ shall be defined by $\coprod \limits_{i \in I} f_i(i, x) = (i, f_i(x))$. The function $\coprod \limits_{i \in I} f_i$ is continuous. Analogously, $\prod \limits_{i \in I} f_i: \prod \limits_{i \in I} (X_i, \mathcal{T}_i) \to \prod \limits_{i \in I} (Y_i, \mathcal{S}_i)$ is defined through $\prod \limits_{i \in I} f_i( \prod \limits_{i \in I} x_i) = \prod \limits_{i \in I} f_i(x_i)$. As abbreviation, $f_1 \pi f_2$ stands for $\prod \limits_{i \in \{1, 2\}} f_i$. For continuous functions $f: (X, \mathcal{T}) \to (Y, \mathcal{S})$ and $g: (Y, \mathcal{S}) \to (Z, \mathcal{R})$, the concatenation  $g \circ f: (X, \mathcal{T}) \to (Z, \mathcal{R})$ is continuous.

As the specific topologies are not relevant for the rest of the paper, we will use the notation $\underline{X}$ to indicate that a set $X$ is equipped with a certain topology. Subsets are equipped with the restriction of the topology of the superset and (co)products of sets with the (co)product topology. A standard reference on topology is the book \cite{dugundji}.

\subsection{Order and Lattice Theory}
A preorder on a class is a binary relation $\preceq$ that is reflexive and transitive. Each preorder defines an equivalence relation $\cong$ via $a \cong b \leftrightarrow a \preceq b \wedge b \preceq a$. On the equivalence classes regarding $\cong$, $\preceq$ becomes a partial order, as it is antisymmetric. In the following, we will not distinguish between a preorder and the partial order on its equivalence classes, the interpretation will be clear from the context.

A partial ordered class $(\mathcal{P}, \preceq)$ is said to be an $\alpha$-complete join-semilattice for a cardinal $\alpha$, if for each $P \subseteq \mathcal{P}$ with $|P| < \alpha$ there is an element $\sup P \in \mathcal{P}$ so that $x \preceq \sup P$ holds for all $x \in P$, and where $\forall y \in P \ y \preceq z$ implies $\sup P \preceq z$. The dual notion is an $\alpha$-complete meet-semilattice, where the existence of $\inf P$ with $\inf P \preceq x$ for $x \in P$ is required, so that $\forall y \in P \ z \preceq y$ implies $z \preceq \inf P$.

A partial ordered class that is an $\alpha$-complete join-semilattice for all cardinals $\alpha$, is called a complete join-semilattice. A partial ordered class that is both an $\alpha$-complete join-semilattice and an $\alpha$-complete meet-semilattice is called an $\alpha$-complete lattice. The definition of a complete lattice is straightforward.

If $\mathcal{Q}$ is a subclass of $\mathcal{P}$, then $(\mathcal{Q}, \preceq)$ is called a sub-join-semilattice of $(\mathcal{P}, \preceq)$, if $\sup P \in \mathcal{Q}$ holds for all $P \subseteq \mathcal{Q}$ .

By choosing $P = \emptyset$, each $\alpha$-complete join-semilattice has a least element $\sup \emptyset$, and every $\alpha$-complete meet-semilattice has a maximal element $\inf \emptyset$. Note that not all results on partial ordered sets are valid for proper classes. An important distinction is that a complete join-semilattice that is defined over a set is also a complete lattice, while this it not necessarily true for an underlying proper class, as $\sup \mathcal{P} = \inf \emptyset$ does not need to exist.

Among the realm of further interesting properties of (semi)lattices is distributivity. While distributivity in the common sense is only defined for lattices, there are several possible ways of extending distributivity to semilattices. We will call a complete join-semilattice distributive, iff $x \leq \sup \limits_{i \in I} y_i$ implies the existence of a family $(x_i)_{i \in I}$ satisfying $x_i \leq y_i$ for all $i \in I$ and $x = \sup \limits_{i \in I} x_i$. For complete lattices, distributivity as defined above is equivalent to the more familiar equation $\inf \{x, \sup \limits_{i \in I} y_i\} = \sup \limits_{i \in I} (\inf \{x, y_i\})$.

A treatment on lattices over sets can be found in \cite{burris} or \cite{graetzer}.

\subsection{Partial Functions and Problems}
In the following, it will be convenient to use partial functions. A partial function  $f: \subseteq X \to Y$ is a function $f: Z \to Y$ with $Z \subseteq X$. A partial function $f: \subseteq \underline{X} \to \underline{Y}$ will be called continuous, if $f: \underline{Z} \to \underline{Y}$ is continuous. A statement such as $f(x) = g(x)$ for partial functions means that either both sides are undefined, or equal.

For some applications, functions are not necessarily an adequate formalisation for the notion of a problem to be solved. In some cases, a problem can be represented by a binary relation linking instances with solutions. We will employ an even more general notion, defining a problem $P: \underline{X} \to \underline{Y}$ to be a set of partial functions from $\underline{X}$ to $\underline{Y}$. A function can be identified with the singleton set containing it, a relation will be identified with the set of its choice functions. We will consider both the problem $\emptyset$ and the problem $\{\emptyset\}$ as relations, the latter being the set containing only the nowhere defined function. The notion of problems was taken from \cite{weihrauchd}.

\subsection{Strongly zero-dimensional metrisable spaces}
For applying the results of the present paper to computable analysis, the topological spaces of particular importance are the strongly zero-dimensional metrisable spaces. The most important examples for this class are the spaces $\underline{\alpha^\mathbb{N}}$ for a cardinal number $\alpha$. The set $\alpha^\mathbb{N}$ is defined as $\alpha^\mathbb{N} = \{f: \mathbb{N} \to \alpha\}$, with the topology derived from the metric $d(f, g) = 2^{-\min\{n \in \mathbb{N} \mid f(n) \neq g(n)\}}$. Of particular relevance is $\underline{\mathbb{N}}^\mathbb{N}$ as it serves as foundation for the theory of representations. A representation of a set $X$ is defined as a surjective partial function $\delta : \subseteq \mathbb{N}^\mathbb{N} \to X$.

We will now define a strongly zero-dimensional metrisable space as a topological space that admits a metric $d$, so that the range of $d$ is $\{0\} \cup \{2^{-n} \mid n \in \mathbb{N}\}$. Clearly, each space $\underline{\alpha^\mathbb{N}}$ is a strongly zero-dimensional metrisable space. On the other hand, each strongly zero-dimensional metrisable space with weight $\alpha$ is homeomorphic to a subspace of $\underline{\alpha^\mathbb{N}}$.

Subspaces, coproducts and countable products of strongly zero-dimensional metrisable spaces are strongly zero-dimensional metrisable spaces. For each $\alpha$, all coproducts of not more than $\alpha$ subspaces of $\underline{\alpha^\mathbb{N}}$ are homeomorphic to a subspace of $\underline{\alpha^\mathbb{N}}$, the same holds for countable products. The results in this subsection are due to \cite{hertling} and \cite{engelking}.

\section{Definitions}
A function $f$ is many-one reducible to a function $g$, if there is a computable function $G$ with $f = g \circ G$. Analogously, $\leq_0$ reducibility is defined using continuous functions. Clearly, the codomain of all functions to be compared with $\leq_0$ has to be fixed.

\begin{definition}
\label{def0basic}
Let $f: \underline{X} \to \underline{Z}$ and $g: \underline{Y} \to \underline{Z}$ be functions. Then $f \leq_0 g$ holds, if there is a continuous function $G: \underline{X} \to \underline{Y}$ with $f = g \circ G$.
\end{definition}

In computable analysis $\leq_0$ is primarily not used for comparing discontinuous functions, but for comparing continuous functions. Especially when studying representations, $\leq_0$ can be interpreted as translatability (\cite{weihrauchb}, \cite{weihrauchd}, \cite{schroder}). Some results for comparing discontinuous functions with $\leq_0$ can be found in \cite{hertling}.

The version of bounded Turing reducibility that is analogous to $\leq_2$ reducibility states that $f$ is reducible to $g$, if $f$ can be computed using one oracle call to $g$. To replace oracle calls with concatenation of functions, the continuous function $\Delta_{\underline{X}}: \underline{X} \to (\underline{X} \times \underline{X})$ defined through $\Delta_{\underline{X}}(x) = (x, x)$ has to be introduced for topological spaces $\underline{X}$. Furthermore, the identity on a topological space $\underline{X}$ is denoted by $id_{\underline{X}}$.

\begin{definition}
\label{def2basic}
Let $f: \underline{X}_1 \to \underline{Y}_1$ and $g: \underline{X}_2 \to \underline{Y}_2$ be functions. Then $f \leq_2 g$ holds, iff there are continuous partial functions $F: \subseteq \underline{X}_1 \times \underline{Y}_2 \to \underline{Y}_1$, $G: \subseteq \underline{X}_1 \to \underline{X}_2$ with $f = F \circ (id_{\underline{X}_1} \pi (g \circ G)) \circ \Delta_{\underline{X}_1}$.
\end{definition}

Note that in Definition \ref{def2basic}, $G$ could also be required to be a function, while requiring $F$ to be a function leads to a different reducibility, as pointed out in \cite[Subsection 1.6.3]{paulymaster}, using an example from \cite[Theorem 2.5.5]{hertling}.

The Definitions \ref{def0basic}, \ref{def2basic} are often restricted through placing certain conditions on the occurring topological spaces. For $\leq_2$, \cite{weihrauchd}, \cite{stein}, \cite{mylatz}, \cite{mylatzb} only consider subspaces\footnote{The consideration of subspaces is hidden in the use of partial functions.} of certain products of $\underline{\mathbb{N}}$ and $\underline{\mathbb{N}^\mathbb{N}}$ or equivalent spaces, \cite{paulymaster} restricts considerations to metric spaces, while \cite{brattka} studies computable metric spaces. \cite{hertling} presents some results for $\leq_2$ restricted to functions with strongly zero-dimensional metrisable spaces as domain and discrete codomain.

While any restrictions on the kind of topological spaces to be considered can be employed for $\leq_0$, as the definition of $\leq_2$ contains some products of the involved spaces, as well as partial functions, it seems reasonable to restrict $\leq_2$ only to classes of topological spaces that are closed under formation of binary products and subspaces.

An extension of \ref{def2basic} to problems is presented in \cite{weihrauchd}, the same approach can also be used for extending $\leq_0$ to problems. The uniform approach employed here, as the functions $F$, $G$ in the following definitions do not depend on $g$, is justified by the interpretation of problems as sets of possible solutions.

\begin{definition}
\label{def0set}
Let $P: \underline{X} \to \underline{Z}$ and $Q: \underline{Y} \to \underline{Z}$ be problems. Define $P \leq_0 Q$, if there is a continuous partial function $G: \underline{X} \to \underline{Y}$ satisfying $g \circ G \in P$ for all $g \in Q$.
\end{definition}

\begin{definition}
\label{def2set}
Let $P: \underline{X}_1 \to \underline{Y}_1$ and $Q: \underline{X}_2 \to \underline{Y}_2$ be problems. Define $P \leq_2 Q$, if there are continuous partial functions $F$, $G$ with $F \circ (id_{\underline{X}_1} \pi (g \circ G)) \circ \Delta_{\underline{X}_1} \in P$ for all $g \in G$.
\end{definition}

It is easy to see that Definitions \ref{def0set}, \ref{def2set} extend the Definitions \ref{def0basic}, \ref{def2basic} when functions are identified with the singleton set containing them. Note especially, that while $G$ was required to be a continuous function in Definition \ref{def0basic}, but a continuous partial function in Definition \ref{def0set}, in the case of singleton sets of functions, $G$ turns to be a function even in Definition \ref{def0set}.

There are further variants of $\leq_2$ that are not restrictions of Definition \ref{def2set}, such as the realizer reducibility introduced in \cite{brattka} or the reducibility for multi-valued functions generalizing realizer reducibility on represented metric spaces from \cite{gherardi}. While it might be possible to transfer our results to these notions, doing so is not within the scope of the present paper.

In the following, we will study equivalence classes for both $\leq_0$ and $\leq_2$. The class of equivalence classes of functions regarding $\leq_i$ is denoted by $\mathbb{F}_i$, the class of equivalence classes of relations by $\mathbb{R}_i$ and the class of equivalence classes for problems by $\mathbb{P}_i$ for $i \in \{0, 2\}$. Note that despite not having been defined explicitly, the reducibilities for relations are obtained as restrictions of the reducibilities for problems.
\section{The induced partial ordered classes}
\subsection{Suprema for $\leq_2$}
Since every preorder induces a partial order on its equivalence classes, in particular $(\mathbb{F}_2, \leq_2)$ is a partial ordered class. As will be proven below, it is even a complete join-semilattice.

\begin{definition}
\label{defsuprema2}
Let $(f_i: \underline{X}_i, \to \underline{Y}_i)_{i \in I}$ be a set-indexed family of functions between topological spaces. Define $\lceil f_i \rceil_{i \in I}: \coprod \underline{X}_i \to \coprod \underline{Y}_i$ through $\lceil f_i \rceil_{i \in I}(i, x) = (i, f(x))$.
\end{definition}

\begin{theorem}
\label{supremum2a}
For all $j \in I$, $f_j \leq_2 \lceil f_i \rceil_{i \in I}$.
\begin{proof}
Define $G(x) = (j, x)$ and $F(x, i, y) = y$. Both functions are continuous w.r.t. the relevant topologies.
\end{proof}
\end{theorem}

\begin{theorem}
\label{supremum2b}
$f_i \leq_2 g$ for all $i \in I$ implies $\lceil f_i \rceil_{i \in I} \leq_2 g$.
\begin{proof}
$f_i \leq_2 g$ implies the existence of suitably defined continuous functions $F_i$, $G_i$ with $f_i(x) = F_i(x, g(G_i(x)))$. Define $F$ through $F(i, x, y) = F_i(x, y)$ and $G$ through $G(i, x) = G_i(x)$. The properties of the coproduct of topological spaces ensure that $F$ and $G$ are continuous w.r.t. the relevant topologies.
\end{proof}
\end{theorem}

\begin{theorem}
\label{supremum2}
$(\mathbb{F}_2, \leq_2)$ is a complete join-semilattice. The suprema are given by $\sup S = \lceil f \rceil_{f \in S}$.
\end{theorem}

Theorem \ref{supremum2} can be transferred to restrictions of $\leq_2$ to suitable classes of topological spaces, as long as these are closed under formation of coproducts. While not all natural examples are closed under arbitrary coproducts, the following theorem provides results for almost all studied restrictions:

\begin{theorem}
\label{supremum2restr}
The partial order induced by the restriction of $\leq_2$ to a class of topological spaces that is closed under formation of $\alpha$-coproducts, is an $\alpha$-complete join-semilattice.
\end{theorem}

Starting from the definition of $\lceil \ \rceil$ for functions, a definition of $\lceil \ \rceil$ for problems can obtained. A separate definition for relations will not be given, but can be obtained as a special case of the following.

\begin{definition}
\label{defsuprema2set}
Let $(P_i: \underline{X}_i \to \underline{Y}_i)_{i \in I}$ be a set-indexed family of problems. Define $\lceil P_i \rceil_{i \in I}$ as $\{\lceil f_i \rceil_{i \in I} \mid \forall i \in I \ f_i \in P_i\}$.
\end{definition}

\begin{theorem}
\label{supremum2aset}
For all $j \in I$, $P_j \leq_2 \lceil P_i \rceil_{i \in I}$.
\begin{proof}
Define $G(x) = (j, x)$ and $F(x, i, y) = y$. Both functions are continuous w.r.t. the relevant topologies. For each $\lceil f_i \rceil_{i \in I} \in \lceil P_i \rceil_{i \in I}$, $F(x, \lceil f_i \rceil_{i \in I}(G(x))) = f_j(x)$ and $f_j \in P_j$ hold, proving the statement.
\end{proof}
\end{theorem}

\begin{theorem}
\label{supremum2bset}
$P_i \leq_2 Q$ for all $i \in I$ implies $\lceil P_i \rceil_{i \in I} \leq_2 Q$.
\begin{proof}
$P_i \leq_2 Q$ implies the existence of suitably defined continuous functions $F_i$, $G_i$ with $x \mapsto F_i(x, g(G_i(x))) \in P_i$ for all $g \in Q$. Define $F$ through $F(i, x, y) = F_i(x, y)$ and $G$ through $G(i, x) = G_i(x)$. The properties of the coproduct ensure that $F$ and $G$ are continuous w.r.t. the relevant topologies. $x \mapsto F(i, x, g(G(i, x)))$ for any $g \in Q$ and fixed $i \in I$ is in $P_i$, so $(i, x) \mapsto F(i, x, g(G(i, x)))$ is in $\lceil P_i \rceil_{i \in I}$.
\end{proof}
\end{theorem}

\begin{theorem}
\label{supremum2set}
$(\mathbb{P}_2, \leq_2)$ is a complete join-semilattice. The suprema are given by $\sup S = \lceil P \rceil_{P \in S}$.
\end{theorem}

Theorem \ref{supremum2restr} holds for relations and problems as well as for functions. Through identifying a function $f$ with the problem $\{f\}$, the partial ordered class $(\mathbb{F}_2, \leq_2)$ is a substructure of the partial ordered class $(\mathbb{P}_2, \leq_2)$. As suprema are formed in a compatible fashion, the complete join-semilattice $(\mathbb{F}_2, \leq_2)$ is even a sub-join-semilattice of $(\mathbb{R}_2, \leq_2)$, and $(\mathbb{R}_2, \leq_2)$ is a sub-join-semilattice of $(\mathbb{P}_2, \leq_2)$.

As the coproduct of an empty family of topological spaces is the space $(\emptyset, \{\emptyset\})$, the minimal element in $(\mathbb{F}_2, \leq_2)$ is the equivalence class containing exactly the functions with domain $\emptyset$. The minimal element in $(\mathbb{P}_2, \leq_2)$ is the equivalence class containing all problems that contain a function with domain $\emptyset$. The continuous functions with non-empty domain form the second-least element of $(\mathbb{F}_2, \leq_2)$, the problems containing a continuous function with non-empty domain are the second-least element of $(\mathbb{P}_2, \leq_2)$.

The restriction to sequential topological spaces even yields a third-least equivalence class of functions containing the function $cf: \underline{\mathbb{N}}^\mathbb{N} \to \{0, 1\}$ with $cf^{-1}(\{1\}) = \{0^\mathbb{N}\}$. This can be rephrased to yield a definition of sequential topological spaces: A topological space $\underline{X}$ is sequential, iff $cf \leq_2 f$ holds for all discontinuous functions $f$ with domain $\underline{X}$. However, even if one restricts problems to those with domain $\underline{\mathbb{N}}^\mathbb{N}$, there exists a decreasing chain between the continuous problems and $\{cf\}$, as shown in \cite[Section 4]{weihrauchc}.

For $(\mathbb{P}_2, \leq_2)$, there exists a maximal element, this contains all empty problems. For functions, however, no maximal element exists, proving that $(\mathbb{F}_2, \leq_2)$ is not an $\alpha$-complete meet-semilattice and therefore not an $\alpha$-complete lattice for any $\alpha > 0$. This claim follows from the examples given at the end of Subsection \ref{subseclevelbase} utilizing the concept of Basesize.

The supremum of a family of relations, considered as problems, can be considered as relation itself, so $(\mathbb{R}_2, \leq_2)$ is a complete join-semilattice, as well as a complete sub-join-semilattice of $(\mathbb{P}_2, \leq_2)$. Furthermore, all examples given above for problems can be considered as relations, so the statements made for $(\mathbb{P}_2, \leq_2)$ also hold for $(\mathbb{R}_2, \leq_2)$.

\subsection{Suprema and Infima for $\leq_0$}
Using a very similar construction to Definition \ref{defsuprema2}, suprema can be introduced for all variations of $\leq_0$ studied here. Again, we will start with considering functions only.
\begin{definition}
\label{defsupremum0}
Let $(f_i : \underline{X_i} \to \underline{Z})_{i \in I}$ be a set-indexed family. Define  $\uparrow f_i \uparrow_{i \in I} : \coprod \limits_{i \in I} \underline{X_i} \to \underline{Z}$ via $\uparrow f_i \uparrow_{i \in I}(i, x) = f_i(x)$.
\end{definition}

\begin{theorem}
\label{supremum0a}
For all $j \in I$, $f_j \leq_0 \uparrow f_i \uparrow_{i \in I}$.
\begin{proof}
Choose $G: \underline{X_j} \to \coprod \limits_{i \in I} \underline{X_i}$ define through $G(x) = (j, x)$.
\end{proof}
\end{theorem}

\begin{theorem}
\label{supremum0b}
If $f_j \leq_0 g$ holds for all $j \in I$, then $\uparrow f_i \uparrow_{i \in I} \leq_0 g$ holds.
\begin{proof}
There are continuous functions $G_j$, so that $f_j = g \circ G_j$ holds for each $j \in J$. Define $G$ through $G(i, x) = G_i(x)$. $G$ is continuous, and satisfies $\uparrow f_i \uparrow_{i \in I} = g \circ G$.
\end{proof}
\end{theorem}

For representations, binary suprema\footnote{Implicitly, Weihrauch introduces also countable suprema and infima.}  for $\leq_0$ have already been introduced in \cite{weihrauchd}. Taking into consideration that $\mathbb{N}^\omega$ and $\mathbb{N}^\omega \coprod \mathbb{N}^\omega$ are homeomorphic,  Definition \ref{defsupremum0} extends \cite[Definition 3.3.11]{weihrauchd}, while the Theorems \ref{supremum0a}, \ref{supremum0b} extend \cite[Theorem 3.3.12 1.]{weihrauchd}. As the restriction of $\leq_0$ to functions with domain in a class of topological spaces closed under formation of $\alpha$-coproducts yields an $\alpha$-complete join-semilattice, also countable suprema exist for representations.

By extending \cite[Definition 3.3.7]{weihrauchd}, a definition of binary infima for representations, $(\mathbb{F}_0, \leq_0)$ can shown to be a complete lattice.
\begin{definition}
Let $(f_i : \underline{X} \to \underline{Z})_{i \in I}$ be a set-indexed family of functions. Define $\downarrow f_i \downarrow_{i \in I} : \mathfrak{P} \to \underline{Z}$, where $\mathfrak{P} = \{\prod \limits_{i \in I} x_i \in \prod \limits_{i \in I} \underline{X}_i \mid \forall i, j \in I \ f_i(x_i) = f_j(x_j)\}$ is equipped with the restriction of the usual product topology, through $\downarrow f_i \downarrow_{i \in I}(\prod \limits_{i \in I} x_i) = f_{i_0}(x_{i_0})$ for an arbitrary fixed $i_0 \in I$.
\end{definition}

\begin{theorem}
\label{infimum0a}
For all $j \in I$, $\downarrow f_i \downarrow_{i \in I} \leq_0 f_j$.
\begin{proof}
Choose $G: \mathfrak{P} \to \underline{X}_j$ as the projection to the $j$th entry.
\end{proof}
\end{theorem}

\begin{theorem}
\label{infimum0b}
Let $g: \underline{Y} \to \underline{Z}$ be a function. If $g \leq_0 f_i$ holds for all $i \in I$, then $g \leq_0 \downarrow f_i \downarrow_{i \in I}$ follows.
\begin{proof}
We assume the existence of continuous functions $G_i$, so that $g = f_i \circ G_i$ holds. This implies $f_i(G_i(y)) = f_j(G_j(y))$ for all $i, j \in I$, $y \in Y$. Thus a continuous function $G: \underline{Y} \to \mathfrak{P}$ can be defined via $G(y) = \prod \limits_{i \in I} G_i(y)$. $G$ satisfies $g = \downarrow f_i \downarrow_{i \in I} \circ G$.
\end{proof}
\end{theorem}

\begin{theorem}
$(\mathbb{F}_0, \leq_0)$ is a complete lattice.
\end{theorem}

The definition of suprema can be extended to relations and problems in the usual manner, as exercised below.
\begin{definition}
Let $(P_i: \underline{X}_i \to \underline{Z})_{i \in I}$ be a set-indexed family of problems. Define $\uparrow P_i \uparrow_{i \in I} = \{\uparrow f_i \uparrow_{i \in I} \mid \forall i \in I \ f_i \in P_i\}$.
\end{definition}

\begin{theorem}
For all $j \in I$, $P_j \leq_0 \uparrow P_i \uparrow_{i \in I}$.
\begin{proof}
Choose $G: \underline{X}_j \to \coprod \limits_{i \in I} \underline{X}_i$ define through $G(x) = (j, x)$. Then $\uparrow f_i \uparrow_{i \in I} \circ G = f_j$ holds, so from $\uparrow f_i \uparrow_{i \in I} \in \uparrow P_i \uparrow_{i \in I}$ follows $\uparrow f_i \uparrow_{i \in I} \circ G \in P_j$.
\end{proof}
\end{theorem}

\begin{theorem}
$P_i \leq_0 B$ for all $i \in I$ implies $\uparrow P_i \uparrow_{i \in I} \leq_0 Q$.
\begin{proof}
There are continuous functions $G_j$, so that $g \circ G_j \in P_j$ holds for each $j \in J$ and each $g \in Q$. Define $G$ through $G(i, x) = G_i(x)$. $G$ is continuous, and satisfies $\uparrow g \circ G_j \uparrow_{i \in I} = g \circ G$, and thus $g \circ G \in \uparrow P_i \uparrow_{i \in I}$ for each $g \in Q$.
\end{proof}
\end{theorem}

As $(\mathbb{F}_0, \leq_0)$ is a complete lattice, there is a smallest and greatest element. The smallest element is the inclusion of the empty set in $\underline{Z}$, the greatest element is the identity $id: (Z, \{\emptyset, Z\}) \to \underline{Z}$. Constant functions are equivalent, iff they have the same image, and incomparable otherwise. Each constant function is a second-smallest element.

Considering problems does not change the results from the last paragraph much, the empty problem is even greater than $\{id\}$, but the equivalence class including $\{id\}$ is the unique second-greatest element.

\subsection{Missing Infima}
So far, the existence of infima was proven or refuted only for the reducibilities for functions. In fact, while $\inf \emptyset$ does not exist in $(\mathbb{F}_2, \leq_2)$, binary or other interesting infima still might be possible. A complete answer cannot be given here, however, we will show that the nature of problems is not compatible with infima in a certain way:

\begin{theorem}
Binary infima in $(\mathbb{F}_i, \leq_i)$ are generally not binary infima in $(\mathbb{P}_i, \leq_i)$ for $i \in \{0, 2\}$.
\begin{proof}
Let $f$, $g$ be functions so that neither $f \leq_i g$ nor $g \leq_i f$ holds. Let the function $h$ be an infimum of $f$ and $g$, and assume that $\{h\}$ is an infimum of $\{f\}$ and $\{g\}$. Consider the problem $\{f, g\}$. Since both $\{f, g\} \leq_i \{f\}$ and $\{f, g\} \leq_i \{g\}$ hold, we infer $\{f, g\} \leq_i \{h\}$. This implies an $h' \in \{f, g\}$ with $h' \leq_i h$. W.l.o.g. assume $h' = f$. Then $f \leq_i g$ follows, contradicting the assumption. As there are incomparable functions, either the infima do not exist, or do not coincide.
\end{proof}
\end{theorem}

\subsection{Distributivity}
\begin{theorem}
\label{distributive2basic}
$(\mathbb{F}_2, \leq_2)$ is distributive.
\begin{proof}
We assume functions $f: \underline{X} \to \underline{Y}$, $g_i: \underline{X}_i \to \underline{Y}_i$ for $i \in I$ satisfying $f \leq_2 \lceil g_i \rceil_{i \in I}$. There are continuous partial functions $F: \subseteq \underline{X} \prod (\coprod \limits_{i \in I} \underline{Y}_i) \to \underline{Y}$ and $G: \subseteq \underline{X} \to \coprod \limits_{i \in I} \underline{X}_i$ with $f(x) = F(x, \lceil g_i \rceil_{i \in I}(G(x)))$ for all $x \in \underline{X}$. We can assume that $G$ is a continuous function. If $\underline{I}$ is the set $I$ with the discrete topology, then the function $\rho: \coprod \limits_{i \in I} \underline{X}_i \to \underline{I}$ defined via $\rho(i, x) = i$ is continuous, and so is $\rho \circ G$. The set $O_i = (\rho \circ G)^{-1}(\{i\})$ for $i \in I$ thus is a open and closed subset of $\underline{X}$.

We use $f_i$ to denote the restriction of $f$ to the set $O_i$. As set-inclusions are continuous, each $f_i$ fulfills $f_i \leq_2 f$, implying $\lceil f_i \rceil \leq_2 f$. Suitable restrictions of $F$ and $G$ also yield $f_i \leq_2 g_i$ for all $i \in I$. It remains to prove $f \leq_2 \lceil f_i \rceil_{i \in I}$. If we use the continuous function $\varrho: \underline{X} \prod (\coprod \limits_{i \in I} \underline{O}_i) \to \underline{X}$ define by $\varrho(x, i, y) = y$, the identity $$f(x) = \varrho(x, \lceil f_i \rceil_{i \in I}((\rho \circ G)(x), x))$$ shows the remaining claim.
\end{proof}
\end{theorem}

\begin{theorem}
\label{distributive2rel}
$(\mathbb{R}_2, \leq_2)$ is distributive.
\begin{proof}
To extend the proof of Theorem \ref{distributive2basic} to relations, note the locality in the definition of a choice function of a relation: If there is a partition $\{p_i \mid i \in I\}$ of a set $X$, so that for a certain function $f$, for each $i \in I$ there is a choice function $g_i$ of a relation $R$, so that the restrictions of $f$ and $g_i$ to $p_i$ are equal, $f$ is a choice function of $R$.
\end{proof}
\end{theorem}

\begin{theorem}
\label{distributive0basic}
$(\mathbb{F}_0, \leq_0)$ is distributive.
\begin{proof}
The proof is exactly parallel to the proof of Theorem \ref{distributive2basic}.
\end{proof}
\end{theorem}

\begin{theorem}
\label{distributive0rel}
$(\mathbb{R}_0, \leq_0)$ is distributive.
\begin{proof}
The proof is exactly parallel to the proof of Theorem \ref{distributive2rel}.
\end{proof}
\end{theorem}

\section{Suprema and Characteristic Numbers}
\subsection{Level and Basesize}
\label{subseclevelbase}
An important tool in the study of the discontinuity of functions are certain characteristic numbers that are compatible with $\leq_2$ (and hence with $\leq_0$). Here, two variants of the Level as introduced in \cite{hertling}, as well as Basesize introduced in \cite{paulymaster} will be considered. While both numbers were defined only for functions originally, they can easily be extended to problems.

\begin{definition}
Let $f: \underline{X} \to \underline{Y}$ be a function. For an ordinal number $\alpha$, inductively define the sets $\mathcal{L}_\alpha^1(f) \subseteq X$ via $\mathcal{L}_0^1(f) = X$, $$\mathcal{L}_{\alpha + 1}^1(f) = \{x \in \mathcal{L}_\alpha^1(f) | f_{|\mathcal{L}_\alpha^1(f)} \textnormal{ is discontinuous in } x\}$$ and $\mathcal{L}_\gamma^1(f) = \bigcap \limits_{\alpha < \gamma} \mathcal{L}_\alpha^1(f)$ for a limit ordinal $\gamma$.
\end{definition}

\begin{definition}
Let $f: \underline{X} \to \underline{Y}$ be a function. For an ordinal number $\alpha$, define inductively the sets $\mathcal{L}_\alpha^2(f) \subseteq X$ via $\mathcal{L}_0^2(f) = X$, $$\mathcal{L}_{\alpha + 1}^2(f) = cl(\{x \in \mathcal{L}_\alpha^2(f) | f_{|\mathcal{L}_\alpha^2(f)} \textnormal{ is discontinuous in } x\})$$ and $\mathcal{L}_\gamma^2(f) = cl \bigcap \limits_{\alpha < \gamma} \mathcal{L}_\alpha^2(f)$ for a limit ordinal $\gamma$.
\end{definition}

\begin{definition}
For $f: \underline{X} \to \underline{Y}$ and $x \in X$, define $\lev^i(f, x) = \min \{\alpha \mid x \notin \mathcal{L}_\alpha^i(f)\}$ and $\Lev^i(f) = \min \{\alpha \mid \mathcal{L}_\alpha^i = \emptyset\}$ for $i \in \{1, 2\}$.
\end{definition}

The formulation of statements involving the Level of a function usually is simplified by assuming that a non-existing Level is comparable with the normal $\leq$ relation for ordinal numbers, and is greater than all ordinal numbers. This agreement extends to suprema and minima of suitable classes of ordinal numbers.

\begin{corollary}
$\Lev^i(f) = \sup \{\lev^i(f, x) \mid x \in X\}$.
\end{corollary}

\begin{theorem}
\label{levelandleq2}
If $f \leq_2 g$ holds, $\Lev^i(f) \leq \Lev^i(g)$ follows.
\begin{proof}
This is the statement of \cite[Korollar 2.4.3]{hertling}.
\end{proof}
\end{theorem}

When trying to define the Level of a problem, two main criteria should be employed. First, the Level of a singleton problem should be identical to the Level of the function it contains. Second, the result of Theorem \ref{levelandleq2} should remain valid when functions are replaced by problems. An elegant way\footnote{The validity of Theorem \ref{levelandleq2} gives $\min \{\Lev^i(f) \mid f \in P\}$ as an upper bound for $\Lev^i(P)$, but the two criteria are not sufficient to uniquely determine Definition \ref{deflevelset}.} of reaching both criteria is presented in the following definition.

\begin{definition}
\label{deflevelset}
Let $P$ be a problem. Define $\Lev^i(P) = \min \{\Lev^i(f) \mid f \in P\}$.
\end{definition}

\begin{theorem}
\label{levelandleq2set}
If $P \leq_2 Q$ holds, $\Lev^i(P) \leq \Lev^i(Q)$ follows.
\begin{proof}
If $P \leq_2 Q$ holds, there are continuous functions $F$, $G$ with $x \mapsto F(x, g(G(x)) \in P$ for all $g \in Q$. Choose a special $g \in Q$, so that $\Lev^i(g) = \Lev^i(Q)$ is fulfilled. Clearly, $x \mapsto F(x, g(G(x)) \leq_2 g$ is true, so from Theorem \ref{levelandleq2} results: $$\Lev^i(x \mapsto F(x, g(G(x))) = \Lev^i(g) = \Lev^i(Q)$$ The claim now follows from Definition \ref{deflevelset}.
\end{proof}
\end{theorem}

The third characteristic number to be considered is Basesize. Basesize extends the notion of $k$-continuity explored in \cite{weihrauchc}. Its definition for functions was first presented in \cite{paulymaster}. In contrast to the Level, the Basesize of a function is a cardinal number.

\begin{definition}
Let $f: \underline{X} \to \underline{Y}$ be a function. A partition for $f$ is a partition $p$ of $X$, so that $f_{|U}$ is continuous for all $U \in p$. The Basesize of $f$ is defined as the least cardinality of a partition for $f$ and denoted by $\bas(f)$.
\end{definition}

\begin{theorem}
\label{basandleq2}
For functions $f: \underline{X} \to \underline{Y}$, $g: \underline{U} \to \underline{V}$, $f \leq_2 g$ implies $\bas(f) \leq \bas(g)$.
\begin{proof}
Let $\{A_i \mid i \in I\}$ be a partition for $g$ with minimal cardinality. Let $F$, $G$ be continuous partial functions with $f(x) = F(x, g(G(x)))$ for all $x \in X$. Then  $\{G^{-1}(A_i) \mid i \in I\}$ is a partition of $X$, and as $g \circ G$ is continuous when restricted to $G^{-1}(A_i)$, so is $f$. So $\{G^{-1}(A_i) \mid i \in I\}$ is a partition for $f$.
\end{proof}
\end{theorem}

The two variants of the Level and Basesize are linked with an inequality. All combinations of equality and strict inequality are possible.
\begin{theorem}
\label{levbasinequalities}
$\bas(f) \leq \Lev^1(f) \leq \Lev^2(f)$.
\end{theorem}

When trying to define the Basesize of a problem, both the goals and the method to achieve them are completely analogous to the same task for the Level.

\begin{definition}
\label{defbasset}
For a problem $P$, define $\bas(P) = \min \{\bas(f) \mid f \in P\}$.
\end{definition}

\begin{theorem}
For problems $P$, $Q$, $P \leq_2 Q$ implies $\bas(P) \leq \bas(Q)$.
\begin{proof}
Choose $g \in Q$ with $\bas(g) = \bas(Q)$. There is an $f \in P$ with $f \leq_2 g$, so $\bas(P) \leq \bas(f) \leq \bas(g) = \bas(Q)$ holds.
\end{proof}
\end{theorem}

Clearly, the inequalities in Theorem \ref{levbasinequalities} hold for problems, too.

In the following, examples will be constructed showing that all combinations of Basesize and Level not ruled out by Theorem \ref{levbasinequalities} can occur. This implies the existence of arbitrarily large antichains in $(\mathbb{F}_2, \leq_2)$, and thus in the other five considered partial ordered classes.

Let $\mathcal{N} = \{0\} \cup \{\frac{1}{n} \mid n \in \mathbb{N}\}$. Given an ordinal number $\lambda$, we let $\mathcal{M}_\lambda$ be the set of order-preserving functions from $\lambda$ to $\mathcal{N}$, both with the standard orders. By identifying $\mathcal{M}_\lambda$ as subset of $\mathcal{N}^\lambda$, a topology on $\mathcal{M}_\lambda$ is obtained as restriction of the usual product topology on $\mathcal{N}^\lambda$. For $c \in \mathcal{M}$, let $F(c) \in \lambda + 1$ denote the least element with $c(F(c)) \neq 0$ or $\lambda$ iff no such element exists.

Given further a cardinal number $\beta$ with $\beta \leq \lambda$, we define a function $R_{\lambda\beta}: (\lambda + 1) \to \beta$ using ordinal left division with remainder. $R_{\lambda\beta}(\alpha)$ shall be the uniquely determined ordinal number less than $\beta$, so that there is an ordinal $\zeta$ with $\alpha = \beta \zeta + R_{\lambda\beta}(\alpha)$. The restriction $\beta \leq \lambda$ ensures the surjectivity of $R_{\lambda\beta}$.

Now a function $f_{\lambda\beta}: \underline{\mathcal{M}}_\lambda \to \underline{\beta}$ is defined as $f_{\lambda\beta} = R_{\lambda\beta} \circ F$, where $\underline{\beta}$ is the set $\beta$ equipped with the discrete topology. We claim $\Lev^1(f_{\lambda\beta}) = \lambda$ and $\bas(f_{\lambda\beta}) = \beta$. The first statement follows from the observation that $\mathcal{L}^1_{\alpha}(f_{\lambda\beta}) = \{ c \in \mathcal{M}_{\lambda} \mid F(c) \geq \alpha\}$. $\bas(f_{\lambda\beta}) \leq \beta$ is clear. $\beta \leq \bas(f_{\lambda\beta})$ can be shown by a straightforward but tedious proof, which is omitted.

\subsection{Permutability of characteristic numbers and suprema}
\begin{theorem}
\label{perlev1sup}
$\Lev^1(\lceil f_i \rceil_{i \in I}) = \sup \{\Lev^1(f_i) \mid i \in I\}$
\begin{proof}
Assume that for each $i \in I$ the domain of $f_i$ is $\underline{X}_i$, so the domain of $\lceil f_i \rceil_{i \in I}$ is $\coprod \limits_{i \in I} \underline{X}_i$. As for each $j \in J$, the set $X_j$ is open and closed in $\coprod \limits_{i \in I} \underline{X}_i$, $\lceil f_i \rceil_{i \in I}$ is continuous in $(j, x)$ iff $f_j$ is continuous in $x$. The same is true for all restrictions. Thus, $\mathcal{L}_\alpha^1(\lceil f_i \rceil_{i \in I}) = \bigcup \limits_{i \in I} \{i\} \times \mathcal{L}_\alpha^1(f_i)$ follows. So $\mathcal{L}_\alpha^1(\lceil f_i \rceil_{i \in I}) = \emptyset$ is true, iff $\mathcal{L}_\alpha^1(f_i) = \emptyset$ holds for all $i \in I$.
\end{proof}
\end{theorem}

\begin{theorem}
\label{perlev2sup}
$\Lev^2(\lceil f_i \rceil_{i \in I}) = \sup \{\Lev^2(f_i) \mid i \in I\}$
\begin{proof}
To prove the claim, the proof of Theorem \ref{perlev1sup} needs to be slightly modified. Therefore, note $cl \coprod \limits_{i \in I} U_i = \coprod \limits_{i \in I} cl U_i$.
\end{proof}
\end{theorem}

\begin{theorem}
\label{perbassup}
$\bas(\lceil f_i \rceil_{i \in I}) = \sup \{\bas(f_i) \mid i \in I \}$
\begin{proof}
The combination of Theorem \ref{supremum2a} and Theorem \ref{basandleq2} yields: $$\bas(\lceil f_i \rceil_{i \in I}) \geq \sup \{\bas(f_i) \mid i \in I \}$$ Now assume an index set $J$ with $|J| = \sup \{\bas(f_i) \mid i \in I \}$. For each $i \in I$, there is a subset $J_i$ of $J$, so that there is a partition $\{U_{ij} \mid j \in J_i\}$ for $f_i$. Define $U_{ij} = \emptyset$ for $j \in J \setminus J_i$. A partition for $\lceil f_i \rceil_{i \in I}$ can be obtained as $\{\bigcup \limits_{i \in I} \{i\} \times U_{ij} \mid j \in J\}$, proving the other direction of the equality.
\end{proof}
\end{theorem}

Again, by building on the result for functions presented in the theorems above, the results can also be obtained for problems. Interestingly, the proof is uniform and not dependent on the specific characteristic number used. This can be regarded as further strengthening the definition of Level and Basesize for problems.

\begin{theorem}
Let $\num \in \{\Lev^1, \Lev^2, \bas\}$. Then follows: $$\num(\lceil P_i \rceil_{i \in I}) = \sup \{\num(P_i) \mid i \in I\}$$
\begin{proof}
According to Definition \ref{defsuprema2set}, $\num(\lceil P_i \rceil_{i \in I}) = \num(\{\lceil f_i \rceil_{i \in I} \mid \forall i \in I \ f_i \in P_i\})$. By Definition \ref{deflevelset} or \ref{defbasset} follows: $$\num(\{\lceil f_i \rceil_{i \in I} \mid \forall i \in I \ f_i \in P_i\}) = \min \{\num(\lceil f_i \rceil_{i \in I}) \mid \forall i \in I \ f_i \in P_i\}$$  Applying Theorem \ref{perlev1sup}, \ref{perlev2sup} or \ref{perbassup}, we obtain: $$\min \{\num(\lceil f_i \rceil_{i \in I}) \mid \forall i \in I \ f_i \in P_i\} = \min \{\sup \{\num (f_i) \mid i \in I\} \mid f_i \in P_i\}$$ $\min$ and $\sup$ commute, so in the next step we have: $$\min \{\sup \{\num (f_i) \mid i \in I\} \mid f_i \in P_i\} = \sup \{ \min \{ \num(f_i) \mid f_i \in P_i\} \mid i \in I\}$$ Another application of Definition \ref{deflevelset} or \ref{defbasset} results in: $$\sup \{ \min \{ \num(f_i) \mid f_i \in P_i\} \mid i \in I\} = \sup \{\num(P_i) \mid i \in I\}$$
\end{proof}
\end{theorem}

\section{Additional Observations}
\subsection{A continuous version of truth-table reducibility}
For some applications the limitation of having only one call to the oracle will be too strict, so a continuous version of truth-table reducibility, meaning the possibility of making any finite number of parallel oracle calls, is desirable. The notion of $n$ parallel calls to an oracle $f$ can be replaced by the notion of one call to the oracle $f^n := \prod \limits_{i = 1}^n f$. The extension to any finite number of calls is accomplished by taking the supremum over all $n$, yielding the following definition:
\begin{definition}
For two functions $f$, $g$, let $f \leq_{ct} g$ hold, if $f \leq_2 \lceil g^n \rceil_{n \in \mathbb{N}}$ holds.
\end{definition}

The transitivity of $\leq_{ct}$ is a consequence of $\lceil f^n \rceil_{n \in \mathbb{N}} \equiv_2 (\lceil f^n \rceil_{n \in \mathbb{N}})^m$ for any function $f$ and natural number $m$. This claim follows from the following distributivity law, which generalises \cite[Theorem 2.2.5.5]{paulymaster}.
\begin{theorem}
$f \pi \lceil g_i \rceil_{i \in I} \equiv_2 \lceil f \pi g_i \rceil_{i \in I}$.
\end{theorem}

Obviously, $\lceil \ \rceil$ can also be considered as supremum in the partial ordered set induced by $\leq_{ct}$, yielding yet another complete join-semilattice. Again it is possible to define $\leq_{ct}$ for relations and problems as well. The join-semilattices corresponding to $\leq_{ct}$ are quotients of the respective join-semilattices for $\leq_2$.

\subsection{Decomposing functions}
When a function $f$ is expressed as a supremum of some functions $f_i$, apparently all questions regarding the discontinuity of $f$ can be answered by examining the functions $f_i$. An example for this is the notion of $C_\infty$-continuous functions introduced in \cite{mylatz}, which corresponds to the supremum of the $\Omega_n$-continuous functions.

For functions defined on a strongly zero-dimensional metrisable space whose Level exists and is a countable limit-ordinal, a general procedure to find an expression as a supremum of less discontinuous functions will be given below. We consider the function $f: \underline{X} \to \underline{Y}$, where $\underline{X}$ is assumed to be metrisable and strongly zero-dimensional. We set $\gamma = \Lev^2(f)$, and let $(\gamma_n)_{n \in \mathbb{N}}$ be an arbitrary sequence satisfying $\gamma_n \leq \gamma$ for all $n \in \mathbb{N}$, as well as $\lim \limits_{i \to \infty} \gamma_i = \gamma$. Further, $L_n$ shall denote the set $\mathcal{L}_{\gamma_n}^2(f)$, and $f_n$ the restriction of $f$ to $X \setminus L_n$.

\begin{theorem}
$f \cong_2 \lceil f_n \rceil_{n \in \mathbb{N}}$
\begin{proof}
As each $f_n$ is a restriction of $f$, for all $n \in \mathbb{N}$, directly $f_n \leq_2 f$ can be obtained. Theorem \ref{supremum2b} yields $\lceil f_n \rceil_{n \in \mathbb{N}} \leq_2 f$.

For the other direction, let $d$ be a metric on $X$ that induces its topology. As $\underline{X}$ is strongly zero-dimensional, the range of $d$ can assumed to be $\mathcal{N} = \{0\} \cup \{\frac{1}{n} \mid n \in \mathbb{N}\}$, equipped with the restriction of the usual Euclidean topology on the real field. For a subset $A \subseteq X$, the function $x \mapsto d(x, A)$ is a continuous function from $\underline{X}$ to $\underline{\mathcal{N}}$. The function $L: \underline{X} \to \prod \limits_{n \in \mathbb{N}} \underline{\mathcal{N}}$, defined by $L(x)(n) = d(x, L_n)$ is also continuous. $\prod \limits_{n \in \mathbb{N}} \underline{\mathcal{N}}$ is homeomorphic to $\underline{\{0, 1\}}^\mathbb{N}$ using $\iota$ as homeomorphism, which is defined via $\iota(w)(\langle n, m\rangle) = 1$ iff $w(n) = \frac{1}{m}$.

By definition, each set $L_n$ is closed, so as $\underline{X}$ is metrisable, $d(x, L_n) = 0$ is equivalent to $x \in L_n$. Since $\bigcap \limits_{n \to \infty} L_n = \emptyset$, for each $x \in X$ there is an $n$ with $x \notin L_n$, so there is an $m$ with $d(x, L_n) = \frac{1}{m}$. So for each $x$, the sequence $\iota(L(x))$ contains a 1. The function which takes a sequence $w \in \{0, 1\}^\omega \setminus \{0^\omega\}$ and returns the least number $n$, so that $w_n$ is 1, is computable and thus continuous. The function $\langle n, m\rangle \mapsto n$ is computable and thus continuous. Concatenation of all these functions yields a continuous function $\mathfrak{L}: \underline{X} \to \underline{\mathbb{N}}_d$ which satisfies $x \notin L_{\mathfrak{L}(x)}$.

Each $x \in X$ thus satisfies $x \in \dom(f_{\mathfrak{L}(x)})$. Therefore, $x \mapsto \lceil f_n \rceil_{n \in \mathbb{N}}(\mathfrak{L}(x), x)$ is well-defined. Concatenation with a projection yields $f(x) = pr(\lceil f_n \rceil_{n \in \mathbb{N}}(\mathfrak{L}(x), x))$, and as both $\mathfrak{L}$ and $pr$ are continuous, this shows $f \leq_2 \lceil f_n \rceil_{n \in \mathbb{N}}$.
\end{proof}
\end{theorem}

\subsection{Defining admissibility via suprema of $\leq_0$}
Admissibility is a desirable property of representations which can be considered central to computable analysis. In \cite{schroder}, Schröder extends the definition of admissibility that e.g. can be found in \cite{weihrauchd} to a more general case, yielding the following definition:
\begin{definition}
\label{admissibility1}
A surjective partial function $\delta :\subseteq \underline{\mathbb{N}}^\mathbb{N} \to \underline{X}$ is called admissible, iff it is continuous and $\rho \leq_0 \delta$ for all continuous surjective partial functions $\rho :\subseteq \underline{\mathbb{N}}^\mathbb{N} \to \underline{X}$.
\end{definition}

Note the following two observations. If $f \leq_0 g$ holds, and $f$ is surjective, so is $g$. If $g_i$ is continuous for $i \in I$, so is $\uparrow g_i \uparrow_{i \in I}$. Then admissibility can be rephrased as a maximality statement regarding the partial order\footnote{This was already noted in \cite{schroder}.} $\leq_0$. We use $\mathcal{C}_p(\underline{X}, \underline{Y})$ to denote the set of continuous partial function from $\underline{X}$ to $\underline{Y}$.

\begin{definition}
\label{admissibility2}
A partial function $\delta :\subseteq \underline{\mathbb{N}}^\mathbb{N} \to \underline{X}$ is called admissible, iff  $\delta \cong_0 \ \uparrow \rho \uparrow_{\rho \in \mathcal{C}_p(\underline{\mathbb{N}}^\mathbb{N}, \underline{X})}$ holds.
\end{definition}

While Definition \ref{admissibility2} does not seem to be more useful than Definition \ref{admissibility1} for practical purposes, it does clearly show the order-theoretic nature of admissibility. Also, Definition \ref{admissibility2} invites the following extension:

\begin{definition}
\label{admissibility3}
A partial function $f : \subseteq \underline{Y} \to \underline{X}$ is called admissible, iff  $f \cong_0 \ \uparrow g \uparrow_{g \in \mathcal{C}(\underline{Y}, \underline{X})}$ holds.
\end{definition}

In \cite{schroder} the topological spaces $\underline{X}$ admitting an admissible representation following Definitions \ref{admissibility1} or \ref{admissibility2} were characterized as those $T_0$-spaces with a countable pseudobase. A generalization of the question lies at hand: Given a topological space $\underline{Y}$, for which topological spaces $\underline{X}$ is there an admissible partial function $f : \underline{Y} \to \underline{X}$? We conclude with giving a trivial answer for a certain subcase: For a discrete space $\underline{D}$, there is an admissible (partial) function $f: \underline{D} \to \underline{X}$, iff $|X| \leq |D|$ holds, as admissibility then coincides with surjectivity. As the class of topological spaces where the underlying sets do not exceed a certain cardinality is not cartesian closed, this example can be considered as a demonstration that $\underline{\mathbb{N}}^\mathbb{N}$ is especially suitable as domain for representations.

\subsection{Generalizing $\leq_0$ in Category Theory}
The simple Definition \ref{def0basic} can easily be formulated in the framework of category theory. Given a category $\mathcal{L}$, a subcategory $\mathcal{K}$ of $\mathcal{L}$ and an object $Z \in \mathcal{L}$, a partial order $\leq_0$ can be defined on the class of morphisms in $\mathcal{L}$ with codomain $Z$:

\begin{definition}
\label{def0cat}
For morphisms $u : X \to Z$, $v: Y \to Z$, $u, v \in \mathcal{L}$, let $u \leq_0 v$ hold, iff there is a morphism $G \in \mathcal{K}$ with $v = u \circ G$.
\end{definition}

While it is not necessary that $\mathcal{K}$ includes all objects from $\mathcal{L}$ for Definition \ref{def0cat} to be valid, this requirement certainly makes $\leq_0$ more useful, so it will be adopted in the following. Note that the trivial case $\mathcal{K} = \mathcal{L}$ is a worthwhile object of study on its own, just as $\leq_0$ can be fruitfully used to compare continuous functions only.

For studying suprema for $\leq_0$, we require that $\mathcal{L}$ has arbitrary coproducts and that $\mathcal{K}$ is closed in $\mathcal{L}$ under formation of coproducts. We recall the definition of coproducts in category theory:
\begin{definition}
Given a family $(A_i)_{i \in I}$ of objects in a category $\mathcal{L}$, an object $A$ together with morphisms $\mu_i : A_i \to A$ is called the coproduct of the $(A_i)_{i \in I}$, iff for every family of morphisms $(f_i : A_i \to Z)$ there is a unique morphism $f: A \to Z$ satisfying $f_i = f \circ \mu_i$ for all $i \in I$.
\end{definition}

We claim that this uniquely determined morphism $f$ is the supremum of the morphisms $f_i$. As $\mathcal{K}$ was required to include all objects and to be closed under formation of coproducts, $\mathcal{K}$ includes all morphisms $\mu_i$, proving $f_i \leq_0 f$ for all $i \in I$. If there is a morphisms $g \in \mathcal{L}$ with morphisms $G_i \in \mathcal{K}$ for $i \in I$ satisfying $g = f_i \circ G_i$, then $g = f \circ (\mu_i \circ G_i)$ follows. Thus $f_i \leq_0 g$ for all $i \in I$ implies $f \leq_0 g$, proving $f$ to be the supremum of the $f_i$.

Studying infima will require the existence of arbitrary pullbacks in $\mathcal{L}$, and the closure of $\mathcal{K}$ in $\mathcal{L}$ under formation of pullbacks, albeit in a very strong sense. Again, we start with recalling the definition of pullbacks:
\begin{definition}
Given a family $(f_i : A_i \to Z)_{i \in I}$ of morphisms in $\mathcal{L}$. The pullback of the $f_i$ is a family of morphisms $(p_i: P \to A_i)_{i \in I}$ satisfying $f_i \circ p_i = f_j \circ p_j$ for all $i, j \in I$, so that if $(q_i : Q \to A_I)_{i \in I}$ is another family of morphisms with $f_i \circ q_i = f_j \circ q_j$, there is a unique morphism $\lambda: Q \to P$ with $q_i = p_i \circ \lambda$ for all $i \in I$.
\end{definition}

The infimum of the family $f_i$ in the definition above is given by the morphism $f = f_i \circ p_i$ (which does not depend on $i$), as long as $p_i \in \mathcal{K}$ for all $i \in I$ and $\lambda \in \mathcal{K}$ are fulfilled. $f \leq_0 f_i$ is clear. Suppose $g \leq_0 f_i$ for all $i \in I$, so there are morphisms $G_i$ with $g = f_i \circ G_i$. This implies $f_i \circ G_i = f_j \circ G_j$, so there is a $\lambda$ with $G_i = p_i \circ \lambda$, thus $g = (f_i \circ p_i) \circ \lambda$ holds, establishing $g \leq_0 f$.

Partial ordered classes can easily be expressed as categories. If $(K, \preceq)$ is a partial ordered class, the associated partial-order-category has the elements of $K$ as objects, and contains a unique morphism $u: A \to B$, iff $A \preceq B$ holds. Concatenation of morphisms is defined straight-forward. Infima in the partial ordered class are pullbacks in the partial-order-category, and suprema in the partial ordered class are coproducts in the partial-order category.


\begin{thebibliography}{10}

\bibitem{bishop}
E.~Bishop and D.S. Bridges.
\newblock {\em Constructive Analysis}.
\newblock Springer Verlag, Berlin, Heidelberg, 1985.

\bibitem{brattka}
Vasco Brattka.
\newblock Effective borel measurability and reducibility of functions.
\newblock In {\em International Conference on Computability and Complexity in
  Analysis}. 2003.

\bibitem{burris}
Stanley Burris and H.P. Sankappanavar.
\newblock {\em A Course in Universal Algebra}.
\newblock Springer, 1981.

\bibitem{dugundji}
James Dugundji.
\newblock {\em Topology}.
\newblock Allyn and Bacon Series in Advanced Mathematics. Allyn and Bacon,
  Boston, 6 edition, 1970.

\bibitem{engelking}
R.~Engelking.
\newblock {\em General Topology}.
\newblock Heldermann, Berlin, 1989.

\bibitem{gherardi}
Guido Gherardi and Alberto Marcone.
\newblock How much incomputable is the separable hahn-banach theorem?
\newblock In Vasco Brattka, Ruth Dillhage, Tanja Grubba, and Angela Klutsch,
  editors, {\em Conference on Computability and Complexity in Analysis}, number
  348 in Informatik Berichte, pages 101 -- 117. FernUniversität Hagen, August
  2008.

\bibitem{graetzer}
George Grätzer.
\newblock {\em General Lattice Theory}.
\newblock Birkhäuser, Basel, 2 edition, 1998.

\bibitem{hertling}
Peter Hertling.
\newblock {\em Unstetigkeitsgrade von Funktionen in der effektiven Analysis}.
\newblock PhD thesis, Fernuniversit{\"a}t, Gesamthochschule in Hagen, Oktober
  1996.

\bibitem{mylatz}
Uwe Mylatz.
\newblock {Vergleich unstetiger Funktionen in der Analysis}.
\newblock Diplomarbeit, FernUniversit{\"a}t Hagen,
  Mai 1992.

\bibitem{mylatzb}
Uwe Mylatz.
\newblock {\em Vergleich unstetiger Funktionen : ``Principle of Omniscience'' \
  und Vollst\"andigkeit in der C-Hierarchie}.
\newblock PhD thesis, Fernuniversit{\"a}t Hagen, Mai 2006.

\bibitem{paulymaster}
Arno Pauly.
\newblock Methoden zum Vergleich der Unstetigkeit von Funktionen.
\newblock Master's thesis, FernUniversität Hagen, March 2007.

\bibitem{schroder}
Matthias Schr\"{o}der.
\newblock Extended admissibility.
\newblock {\em Theoretical Computer Science}, 284(2):519--538, 2002.

\bibitem{simpson2}
S.G. Simpson.
\newblock {\em Subsystems of second order arithmetic}.
\newblock Springer, Berlin, 1999.

\bibitem{stein}
Thorsten~von Stein.
\newblock Vergleich nicht konstruktiv l{\"o}sbarer Probleme in der Analysis.
\newblock Diplomarbeit, Fachbereich Informatik, FernUniversit\"at Hagen,
  1989.

\bibitem{troelstra}
A.S. Troelstra.
\newblock Comparing the theory of representations and constructive mathematics.
\newblock In E.~Börger, G.~Jäger, H.~Kleine-Büning, and M.M. Richter, editors,
  {\em Computer Science Logic}, volume 626 of {\em Lecture Notes in Computer
  Science}, pages 382 -- 395. Springer, 1992.

\bibitem{weihrauchb}
Klaus Weihrauch.
\newblock The degrees of discontinuity of some translators between
  representations of the real numbers.
\newblock Informatik Berichte 129, FernUniversit\"at Hagen, Hagen, July 1992.

\bibitem{weihrauchc}
Klaus Weihrauch.
\newblock The {TTE}-interpretation of three hierarchies of omniscience
  principles.
\newblock Informatik Berichte 130, FernUniversit\"at Hagen, Hagen, September
  1992.

\bibitem{weihrauchd}
Klaus Weihrauch.
\newblock {\em Computable Analysis}.
\newblock Springer-Verlag, 2000.

\end{thebibliography}
\end{document}